\documentclass[fleqn,usenatbib]{mnras}

\usepackage[T1]{fontenc}
\usepackage{ae,aecompl} 

\usepackage{graphicx}   % Including figure files
\usepackage{amsmath}    % Advanced maths commands
\usepackage{amssymb}    % Extra maths symbols

\newcommand{\kms}{km\,s$^{-1}$\,}

\newcommand{\Ha}{H${\alpha}$ }

\newcommand{\Ng}{300~}
\newcommand{\Neon}{561~}

\title[Lags in edge-on galaxies]
{SDSS IV MaNGA - Gas Rotation Velocity lags in the Final Sample of MaNGA Galaxies}

\author[Bizyaev et al.]{
Dmitry Bizyaev$^{1,2,3}$\thanks{E-mail: dmbiz@apo.nmsu.edu},
Rene A. M. Walterbos$^{4}$,
Yan-Mei Chen$^{5,6}$,
Niv Drory$^{7}$,
Richard R. Lane$^{8}$,  \\
\newauthor Joel R. Brownstein $^{9}$,
Rogemar A. Riffel $^{10,11}$ \\
$^{1}$Apache Point Observatory and New Mexico State University, Sunspot, NM, 88349\\
$^{2}$Sternberg Astronomical Institute, Moscow State University, Universitetskiy prosp. 13, Moscow, 119234\\
$^{3}$Special Astrophysical Observatory of the Russian AS, 369167, Nizhnij Arkhyz\\
$^{4}$New Mexico State University, Last Cruces, NM, 88003, USA\\
$^{5}$School of Astronomy and Space Science, Nanjing University, Nanjing 210093, China \\
$^{6}$Key Laboratory of Modern Astronomy and Astrophysics (Nanjing University), Ministry of Education, Nanjing 210093, China\\
$^{7}$McDonald Observatory, The University of Texas at Austin, 1 University Station, Austin, TX 78712, USA \\
$^{8}$Centro de Investigac\'ion en Astronom\'ia, Universidad Bernardo
O'Higgins, Avenida Viel 1497, Santiago, Chile \\
$^{9}$Department of Physics and Astronomy, University of Utah, 115 S.
1400 E., Salt Lake City, UT 84112, USA \\
$^{10}$Departamento de F\'isica, CCNE, Universidade Federal de Santa Maria, 97105-900, Santa Maria, RS, Brazil \\
$^{11}$Laborat\'orio Interinstitucional de e-Astronomia - LIneA, Rua Gal. Jos\'e Cristino 77, Rio de Janeiro, RJ - 20921-400, Brazil   \\
}
 
\begin{document}
 
% These dates will be filled out by the publisher
\date{Accepted XXX. Received YYY; in original form ZZZ}

% Enter the current year, for the copyright statements etc.
\pubyear{2022}

\label{firstpage}
\pagerange{\pageref{firstpage}--\pageref{lastpage}}
\maketitle

\begin{abstract}

We consider the largest sample of \Neon edge-on galaxies observed with 
integral field units by the MaNGA survey and 
find \Ng galaxies where the ionised gas
shows a negative vertical gradient (lag) in its rotational speed.
We introduce the stop altitude as the distance to the galactic midplane at which 
the gas rotation should stop in the linear approximation. We find correlations
between the lags, stop altitude and galactic mass, stellar velocity 
dispersion and overall Sersic index. We do not find any correlation of 
the lags or stop altitude with the star formation activity in the galaxies. 
We conclude that low mass galaxies (log(M$_*/M_{\odot}$) < 10) with 
low Sersic index and with low stellar velocity dispersion posses a wider
"zone of influence" in the extragalactic gas surrounding them with respect to 
higher mass galaxies that have a significant spherical component. 
We estimated the trend of the vertical rotational gradient with radius and find it 
flat for most of the galaxies in our sample.
A small subsample of galaxies with negative radial gradients of lag
has an enhanced fraction of objects with aged low surface brightness 
structures around them (e.g. faint shells), which indicates that 
noticeable accretion events in the past affected the extraplanar 
gas kinematics and might have contributed to negative radial lag gradients. 
We conclude that an isotropic accretion
of gas from the circumgalactic medium plays a significant role in the formation of
rotation velocity lags. 

\end{abstract}


%%%%%%%%%%%%%%%%%%%%%%%%%%%%%%%%%%%%%%%%%%%%%%%%%%%%%%%%%%%%%%%%%%%%
\section{Introduction}

Ionized gas is found not only close to the midplane, but also 
can be seen at high altitudes in our Galaxy
\citep{hoyle63,reynolds71,reynolds73,reynolds99,shull09} as well as
in other galaxies
\citep{dettmar90,rand90,rand97,hoopes99,rand00,rossa04} \citep{rossa03,wu14},
at several kpc from the galactic
midplane.  The high-altitude gas plays important role for the 
matter circulation in galaxies \citep{putman12}. It may replenish
the in-plane gas and support star formation process. 

There are different scenarios of the origin of the extra-planar gas. 
Accretion from the external intergalactic medium not connected to the
inner galactic gas is one of them \citep{oort70, white78, binney05, kaufmann06, combes14}. 
Galactic fountains recycle galactic gas and cause its accretion later
\citep{shapiro76,bregman80,norman89,marinacci11,fraternali13,lehner22,marasco22}. 
These effects do not exclude each other and can play
together \citep{haffner09,benjamin12}.  

As it has been noticed for several nearby galaxies, the gas rotation
deviates from cylindrical and sometimes we observe rotation velocity lag.
The lags were noticed in atomic 
\citep{swaters97,matthews03,zschaechner11,gentile13,kamphius13,zschaechner15a,zschaechner15b} and in ionized gas \citep{fraternali04,heald06a,heald06b, heald07,kamphius07,kamphius11,rosado13,wu14,boettcher16}.

A strong correlation between star formation per unit disc area
and the existence of ionized extra planar gas 
has been noticed \citep{dettmar04,ho16}.
Contrary to expectations that lags are connected to the star formation 
activity in galaxies, an inverse
correlation between the star formation rate and the 
lag amplitude was reported by \citet{heald07}.  \citet{zschaechner11} did not
confirm it. A large sample of edge-on galaxies observed by the MaNGA
\citep{bundy15} survey was investigated by \citet{bizyaev17} and no connection between
the star formation and the lags was found. This result was also confirmed by 
\citet{levy19} with CALIFA \citep{califa} survey data. 

Correlation of lags with galactic properties can reveal the nature of
the extraplanar ionized gas. Thus, \citet{zschaechner15a} reported 
negative radial gradients of the lags (i.e. the magnitude of the lag
decreases toward the periphery of the disc), which points towards the galactic
fountains as the main source of the extraplanar gas. Studying HI velocity
fields in nearby galaxies in the frames of the HALOGAS project revealed galactic
fountains as good explanation for the lags in nearby late-type galaxies 
\citep{marasco19}. On the other hand,
\citet{levy19} do not find systematic lags in their sample of galaxies 
and conclude that the main source of the extraplanar gas is an external
accretion. The gas accretion from the CGM is a necessary component for
explaining kinematics of ionized and neutral gas around galaxies 
according to \citet{fraternali07,fraternali08,marinacci10}, even in the presence
of galactic fountains. 

To this date, three large Integral Field Unit surveys have been finished 
(CALIFA\footnote{Calar Alto Legacy Integral Field spectroscopy Area survey}, \citet{califa}, 
MaNGA\footnote{Mapping Nearby Galaxies at Apache Point Observatory}, \citet{bundy15,sdssdr17}, and
SAMI\footnote{Sydney-Anglo-Australian Observatory Multi-object Integral field survey}, \citet{sami}) and released their final data. They provide an excellent 
opportunity for studying the ionized gas in galaxies and around them, as
well as its connection to the environment of galaxies \citep{bloom18}. 
 
In this paper we utilize the data from the MaNGA survey released at the end
of 2021 \citep{sdssdr17}. We concentrate on edge-on galaxies only, which helps minimize
projection effects in the galaxies and select regions populated
with extraplanar gas only. We called the vertical gradient of rotation velocity
as lag throughout the paper.

\section{The Final Data Release of MaNGA}

This year the SDSS\footnote{Sloan Digital Sky Survey, http://sdss.org}
has issued the final release of MaNGA panoramic spectroscopy data
for 10,010 unique galaxies in the frames of SDSS-IV \citep{blanton17}. 
MaNGA is a massive spectroscopic survey
performed with IFU technique \citep{bundy15,drory15}, which was run
on the Sloan 2.5 m telescope at Apache Point Observatory
\citep{gunn06,smee13}. The survey employs a simple, "flat"
target selection by stellar mass of the galaxies \citep{wake17}.
MaNGA uses 17 IFUs simultaneously with size from 12 to 32 arcsec 
(19 to 127 fibers, respectively), see \citet{drory15}.  
The precision of the flux calibration is at the level of a percent
\citep{yan16}, with the best spectral resolution $\sigma \,\sim~$
25 \kms  \citep{law21}. The median redshift of the survey's object 
is about 0.03 \citep{law16}. The final MaNGA data release
provides raw reduced spectra \citep{law16} as well as numerous
high-level products such as maps of gas and stellar kinematics,
spectral indices, fluxes in individual emission lines, and many 
others \citep{westfall19,sdssdr17}.

\subsection{Edge-on Galaxies observed by MaNGA}
Following the approach by \citet[][B17 hereafter]{bizyaev17}, we visually inspected SDSS
images of all MaNGA galaxies. The objects we considered
to be true edge-on galaxies if they did not show spiral arms
and had a dust lane projected very close to the galactic disc 
midplane and nucleus. We were able to identify \Neon edge-on
galaxies. 

\subsection{Edge-on Galaxies with Regular Lags}

Following B17, we plotted radial velocity profiles for
selected galaxies drawn parallel to the minor axis and 
searched for regularly descending velocities with the
distance to midplane. We used the \Ha emission velocity 
maps and only the spaxels with the signal-to-noise ratio 
SNR$>$3 and without bad processing flags set by the data reduction
pipelines. As a result we selected \Ng edge-on galaxies
with regular rotation velocity lags. SDSS images of
the galaxies are shown on Figure~\ref{fig1}.

\clearpage
\begin{figure}
%\epsscale{0.5}
%%%\plotone{f1b.png}
%\plotone{fig2.png}
%\plotone{fig1a.pdf}
\includegraphics[width=\textwidth]{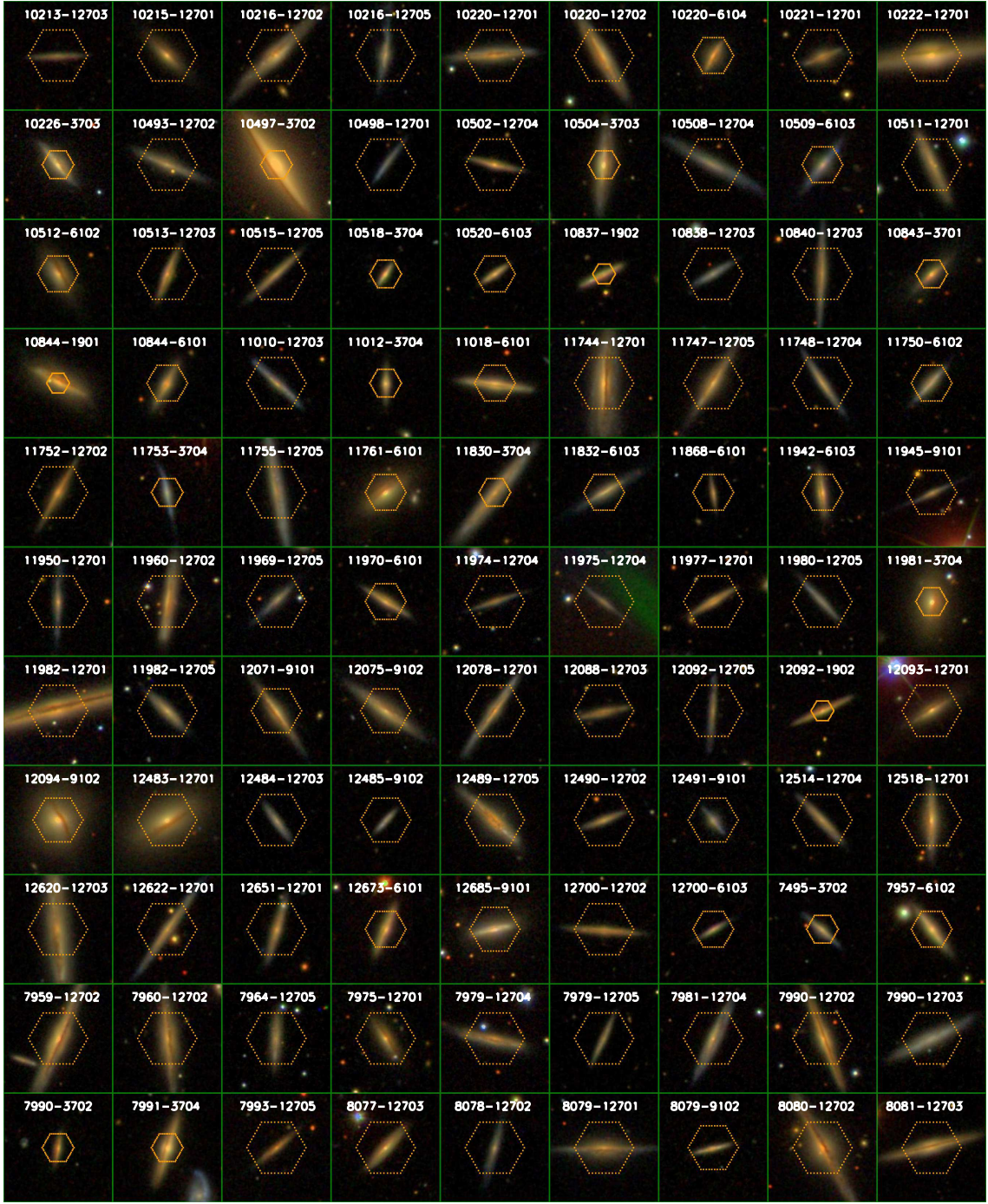}
\caption{SDSS images of selected edge-on galaxies with regular 
rotation curve lags. All images show 1 arcmin x 1 arcmin square
area.
The yellow contours designates the MaNGA IFU allocation. 
\label{fig1}}
\end{figure}
\clearpage

\begin{figure}
%\epsscale{1.0}
%%%\plotone{f1b.png}
%\plotone{fig2.png}
%\plotone{fig1b.pdf}
\includegraphics[width=\textwidth]{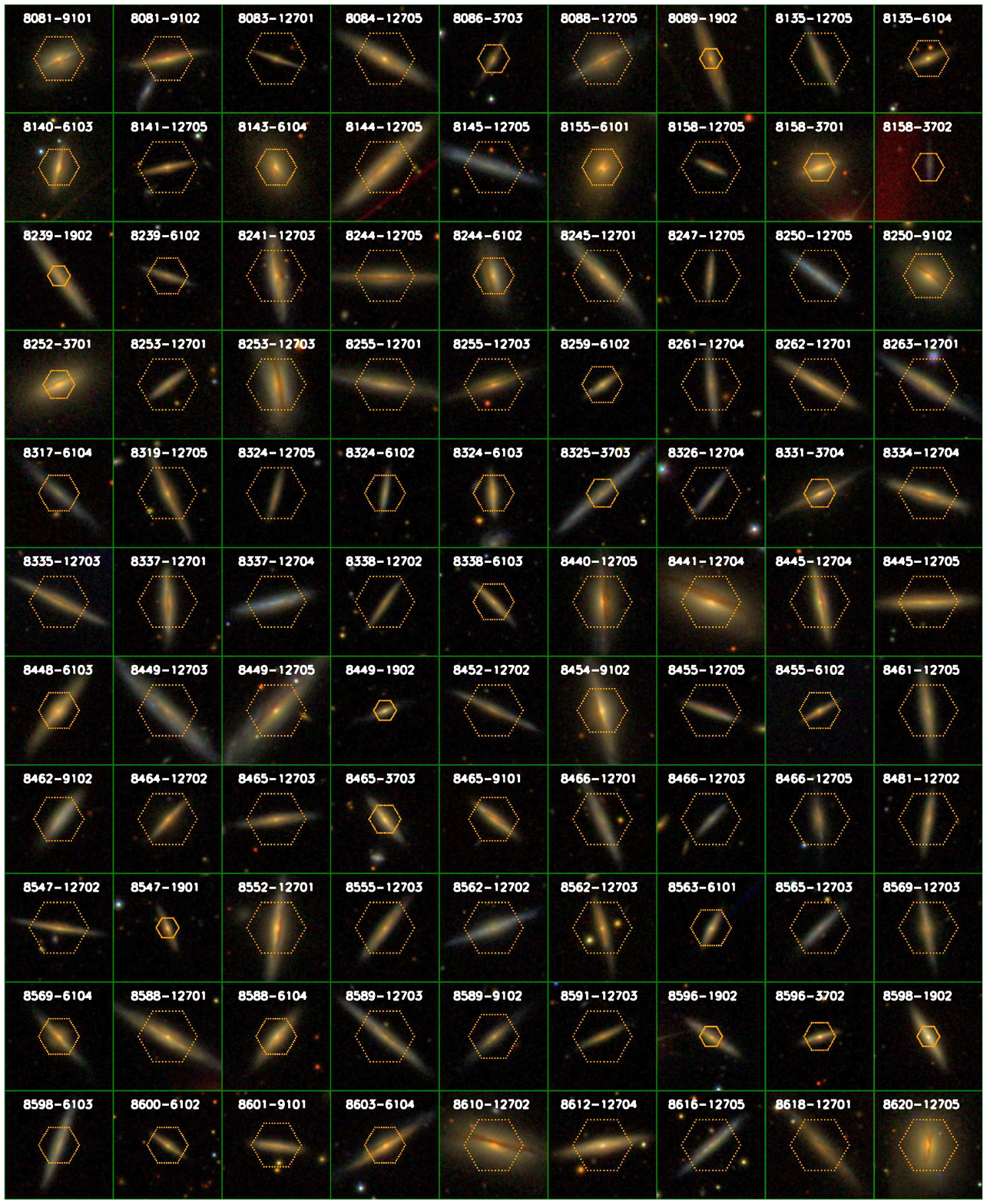}
\setcounter{figure}{0}
\caption{
Continue of Fig. 1
\label{fig1b}}
\end{figure}
\clearpage

\begin{figure}
%\epsscale{1.0}
%%%\plotone{f1b.png}
%\plotone{fig2.png}
%\plotone{fig1c.pdf}
\includegraphics[width=\textwidth]{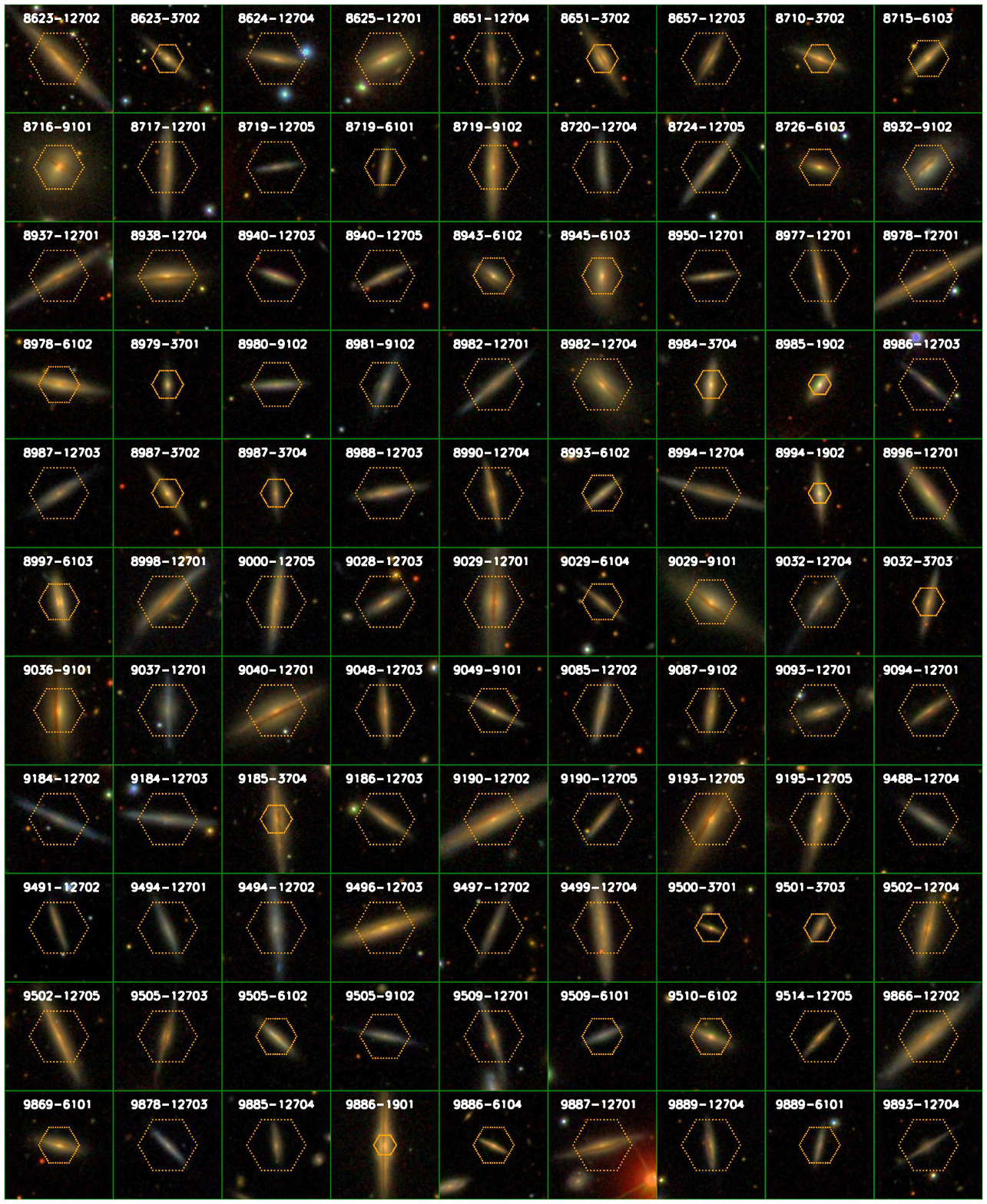}
\setcounter{figure}{0}
\caption{
Continue of Fig. 1
\label{fig1c}}
\end{figure}
\clearpage

\section{Ionized Gas Velocity Fields}
\subsection{Modelling Velocity Fields with Lags}

We use benefits of two-dimensional velocity fields provided by panoramic spectroscopy 
and compare them to the same model used in B17. The model assumes that the \Ha emission 
follows an exponential distribution along the radial and vertical directions in the 
discs. We estimate the radial and vertical scales of for \Ha maps made with the 
MaNGA data cubes and the same procedure as in \citet{bizyaev02,bizyaev14}.
The rotation curved is parameterized via a simple function with a linear 
raising part at the centre, see B17. The three-dimensional distribution of the
velocities is projected on the sky plane by integrating along the line of 
sight. The dust extinction is taken into account in the model via introducing a 
co-planar exponential dust disc. Similar to B17, the dust disc has the same 
spatial scales as the gas disc, and the central extinction of the dust is a 
free parameter of the model. Finally, the projected radial velocity maps are 
convolved with a gaussian kernel corresponding to the spatial resolution 
of MaNGA emission line velocity maps. The convolution uses corresponding \Ha fluxes
as weights. The model parameters are estimated via
the chi-square minimization of the difference between the model maps and observations. 

Same as in B17, the lag of the rotation is introduced as a vertical gradient
of rotation curve: $V(r,z) = V(r) - |z|\, dv/dz$, where $r$ and $z$ are the radial
and vertical coordinates on the sky plane, and $V$ is the projected radial velocity. 
We have shown (B17) that introducing the lag significantly improves the 
projects velocity fitting to the observing data if the estimated lag value 
exceeds a few \kms. 

\begin{figure}
\includegraphics[width=8.5cm]{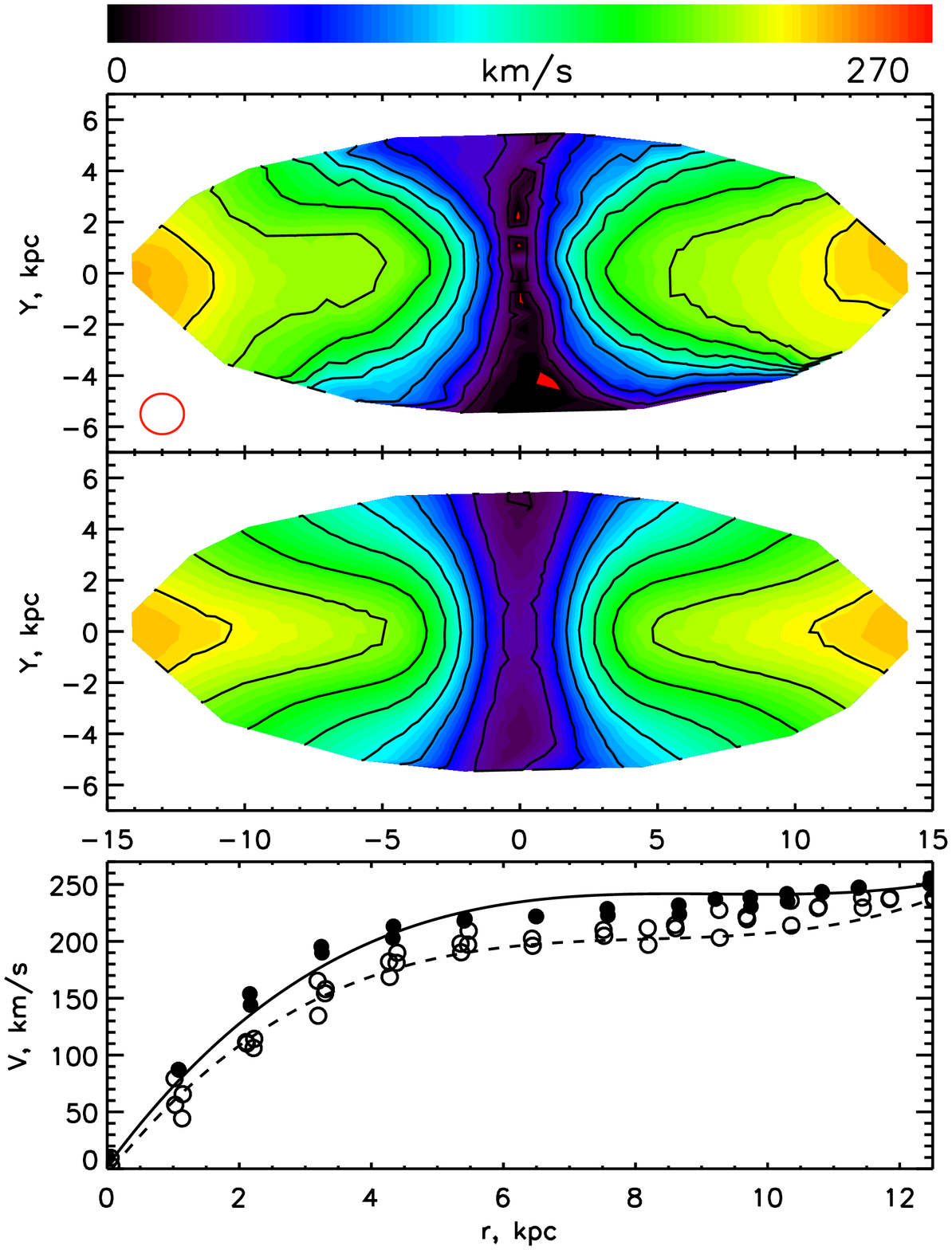}
\caption{An example of the velocity field fitting, the MaNGA galaxy 1-122111.
Top panel: the MaNGA observing velocity field. We show absolute velocities on the plot.
The red ellipse at the lower left corner of the panel indicates the MaNGA spatial resolution (sigma). 
Middle: the model velocity
field. Bottom: two observing (bullets) and model (curves) rotation 
curves in the galactic midplane (filled symbols) and at 2 kpc above
the midplane (open symbols). 
\label{fig1d}}
\end{figure}

Figure~\ref{fig1d} shows results of our fitting for a typical MaNGA edge-on galaxy. 
The observed and model velocity fields are shown along with the observed
and model rotation curves in the galactic midplane and at 2 kpc above it. 
Due to the vertical lag in rotational velocities,
the amplitude of the rotation curve
is lower at high altitude with respect to the
midplane. 

\subsection{Rotation Velocity Lags in the Galaxies}

\begin{figure}
%\epsscale{1.0}
%\plotone{fig2.eps}
%\includegraphics[width=\textwidth]{fig2.eps}
\includegraphics[width=8.5cm]{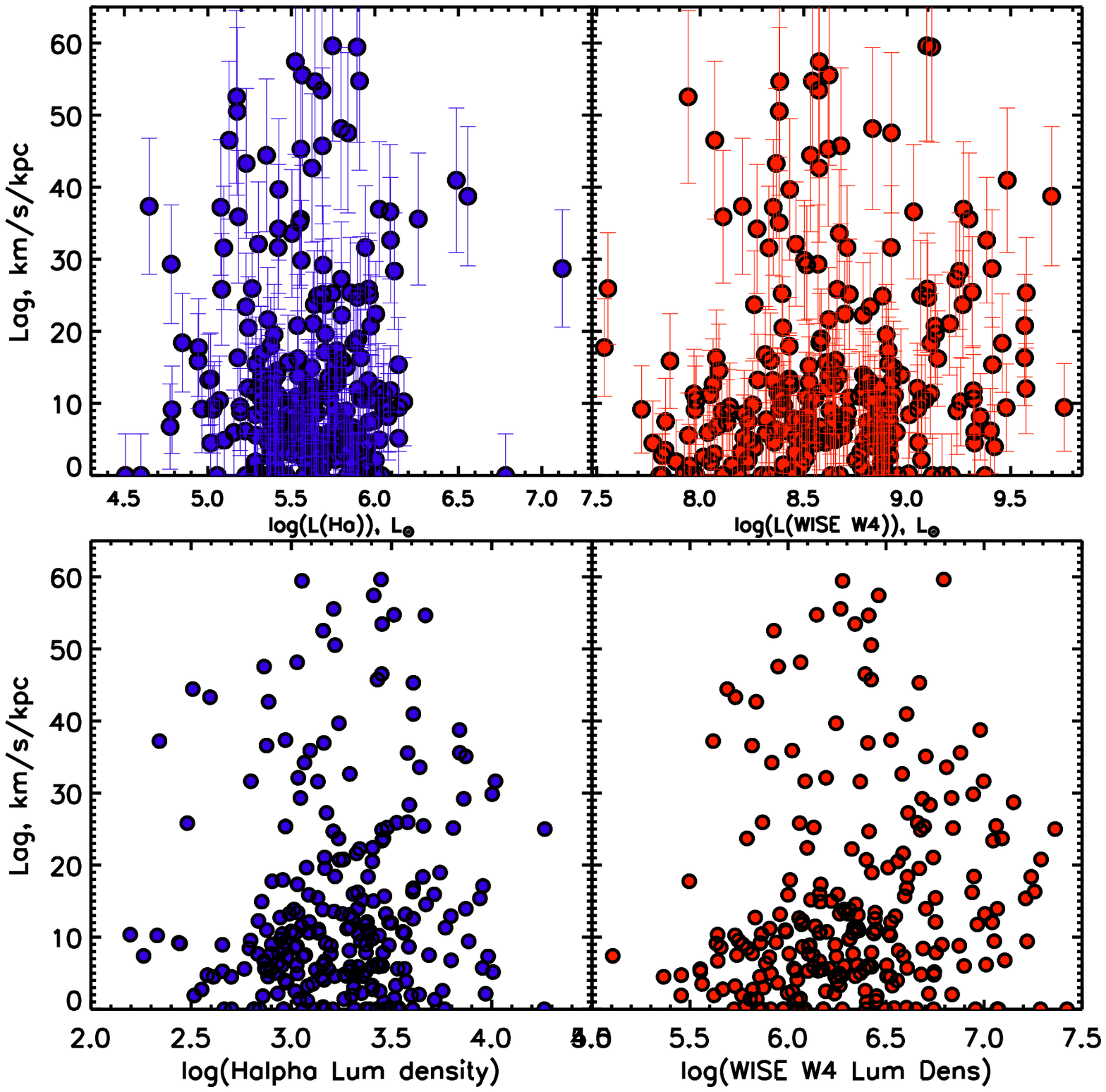}
\caption{The rotation velocity lag versus the star formation properties
in the galaxies: the \Ha and WISE W4 luminosities (upper row)
and their surface density (lower row).
\label{fig2}}
\end{figure}

We observe no correlation with star formation activity in our sample of galaxies.
We applied two approaches to estimate the star formation rate: via \Ha luminosity, and 
via WISE W4 band flux. There is no dependence can be seen in Figure~\ref{fig2}:
relationships on both left and right top panels have the Pearson correlation coefficient
(CC hereafter) of 0.07.
We also considered an average star formation density estimated as the integrated \Ha or W4 
fluxes divided by galactic disc area. The area is estimated as $2\, \pi \, R_{petro}^2$, 
where $R_{petro}$ is the Petrosian disc radius from the parental NASA-Sloan Atlas, 
\citet{blanton11} (NSA\footnote{http://nsatlas.org}) catalog. 
Lags do not show correlation with the star formation density as well: 
the bottom panels in Figure~\ref{fig2} correspond to CC = 0.09 and 0.10 for the left and 
right panels, respectively.

\begin{figure}
\includegraphics[width=8.5cm]{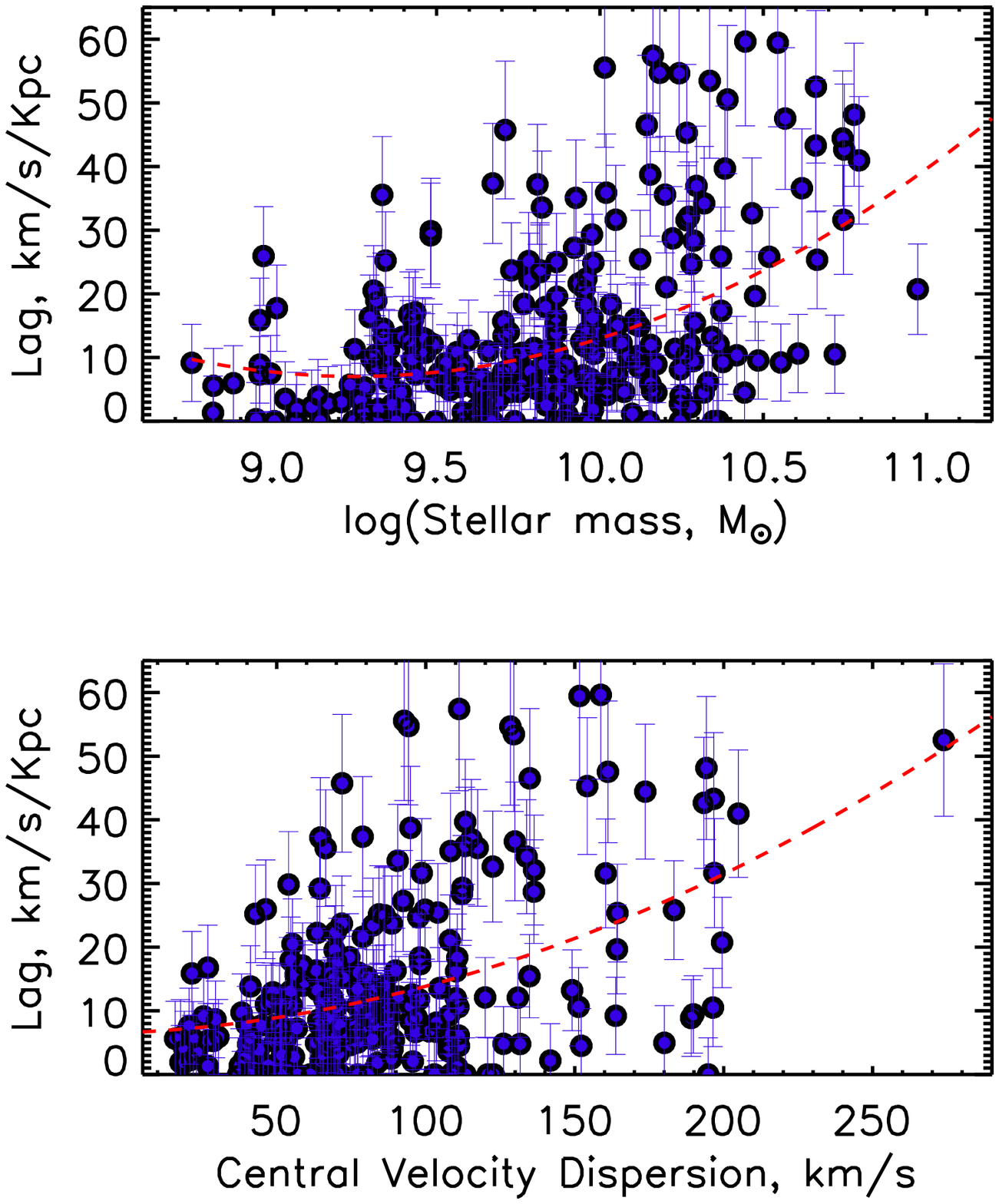}
\caption{The rotation velocity lag versus the stellar mass (upper)
and central stellar velocity dispersion (lower). The red dashed
curve is a polynomial regression of the data. 
\label{fig3}}
\end{figure}

\begin{figure}
\includegraphics[width=8.5cm]{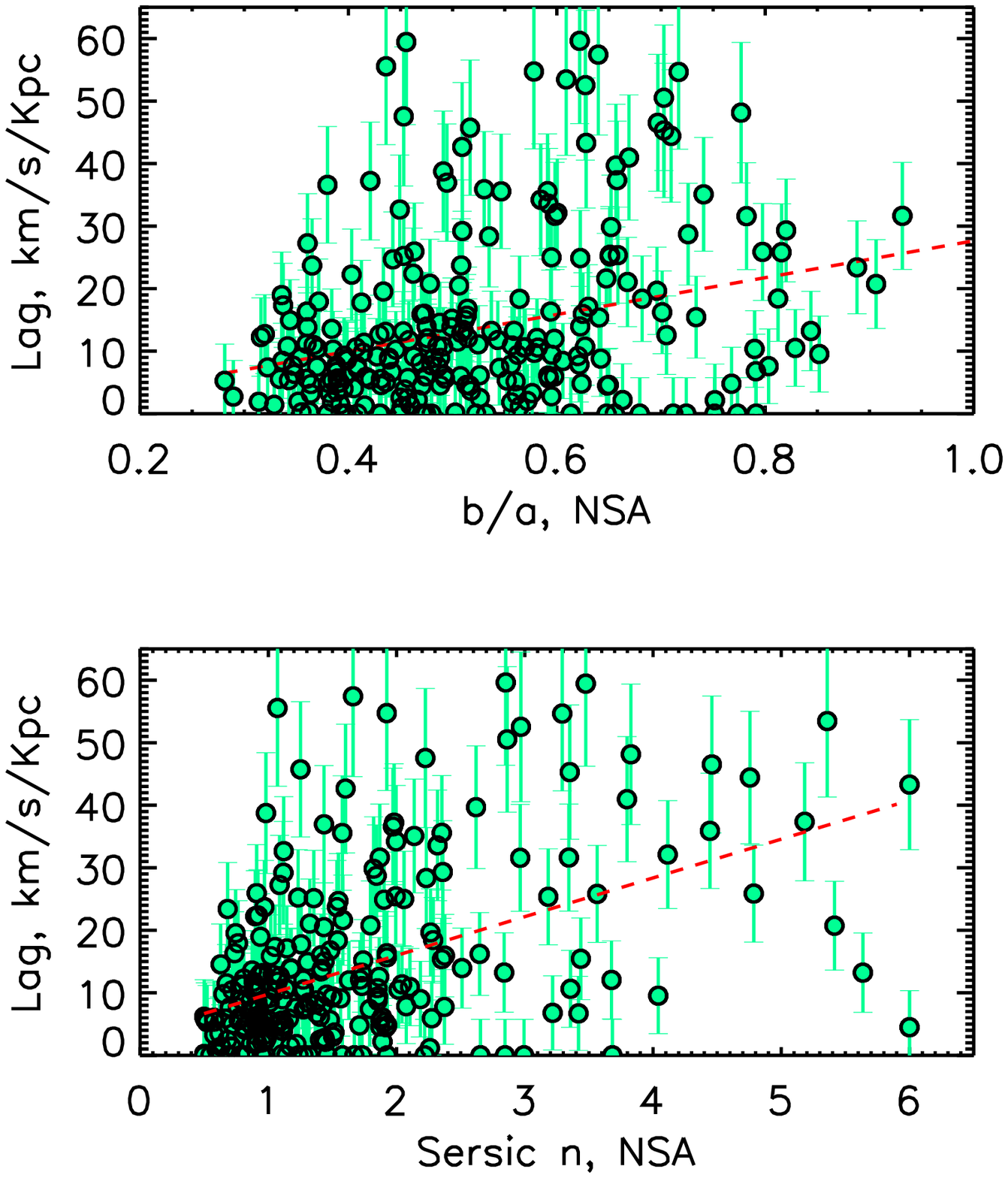}
\caption{The rotation velocity lag versus the apparent ellipticity 
of the edge-on galaxies and their Sersic index taken from 
the NSA catalog. The red dashed curve shows a polynomial regression. 
\label{fig4}}
\end{figure}

Similar to B17, lags show tendency to be higher for galaxies with large
stellar mass and central velocity dispersion, see Figure~\ref{fig3}.
The correlation coefficient for both panels in Figure~\ref{fig3}
is CC = 0.43.
The visual ellipticity of the galaxies, as well as their Sersic index
"n" weakly correlate with the value of lags as well in Figure~\ref{fig4}:
CC = 0.30 for the upper panel and CC = 0.48 for the lower one.

\begin{figure}
\includegraphics[width=8.5cm]{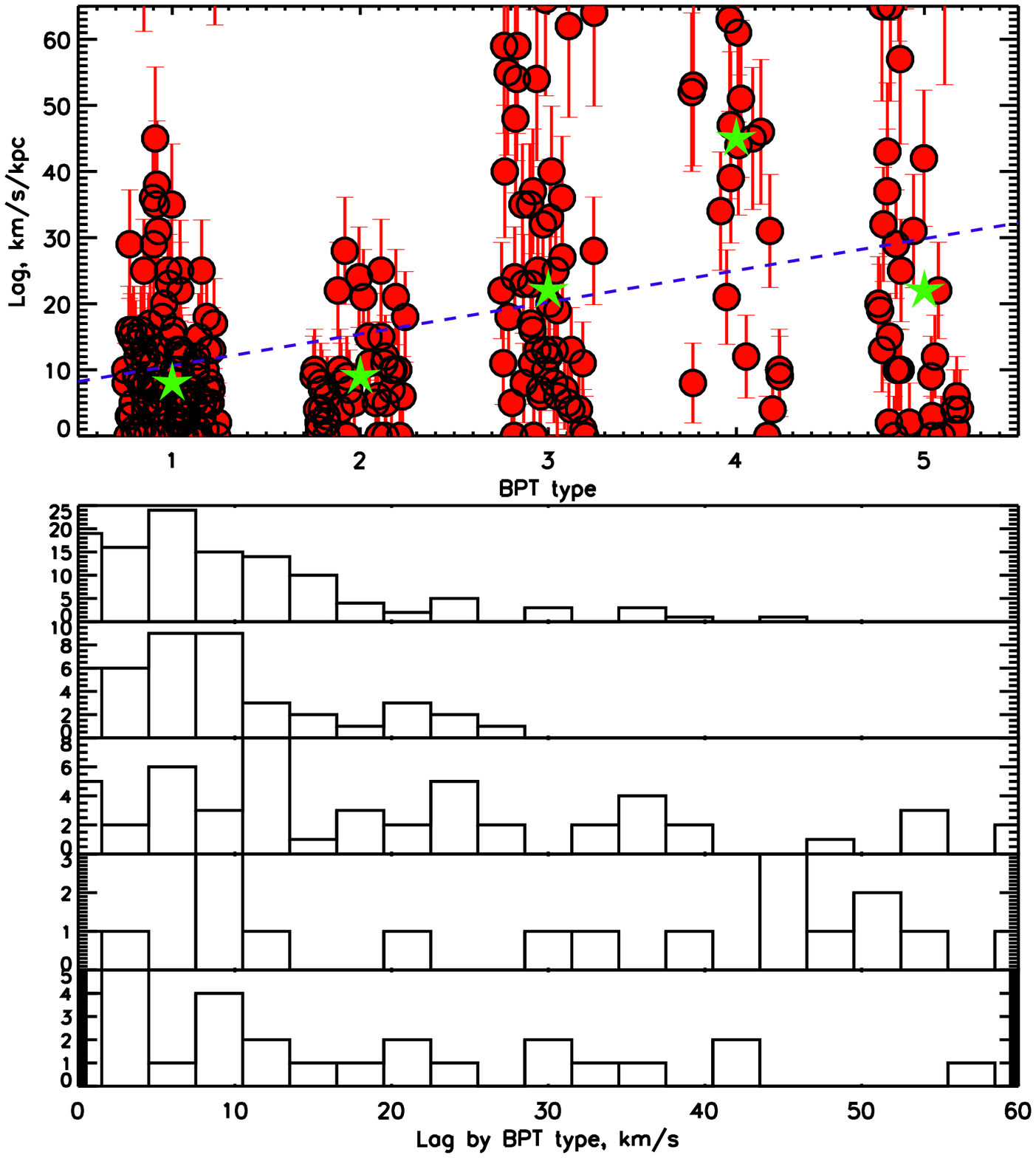}
\caption{Top: the rotation velocity lag is shown for different
BPT classes \citep{brinchmann04}. The classes 1-2 designate
objects with emission line ration from the centre typical 
for star forming regions. The class 3 marks composite regions.
The classes 4 and 5 designate the AGN and LINERs nuclei, 
respectively. The points are randomly scattered in the X-direction
for a better representation. The blue dashed line shows
the linear regression. It demonstrates about three times
larger lags in the AGN-LINER galaxies with respect to the 
star forming ones. The green stars designate the median 
values of the lag in each BPT class. 
Bottom: each BPT class from the top plot is shown
as a separate histogram (class 1 to 5 from top to bottom). 
\label{fig5}}
\end{figure}

The emission lines flux in the central spaxel of the galaxies allows to 
classify them on the BPT diagram \citep{BPT,veilleux87}. 
For consistency with B17, we used the classifications based on SDSS spectroscopic 
survey spectra, which were estimated independently of MaNGA. The data are taken from the MPA-JHU catalog 
\citep{brinchmann04}. The classes 1 and 2 correspond to star forming and
low signal-to-noise star forming regions; class 3 designates composite regions;
class 4 shows the AGN (excluding LINERs) nuclei; class 5 corresponds to LINERs.
Similar to B17, the galaxies with AGNs and LINERs tend to have larger lags,
while galaxies with star-forming nuclei seldom have extraplanar gas with 
high rotation lag. The lag difference between the classes 1-2 and
4-5 is about 3 times, see Figure~\ref{fig5}.

\begin{figure*}
\includegraphics[width=16cm]{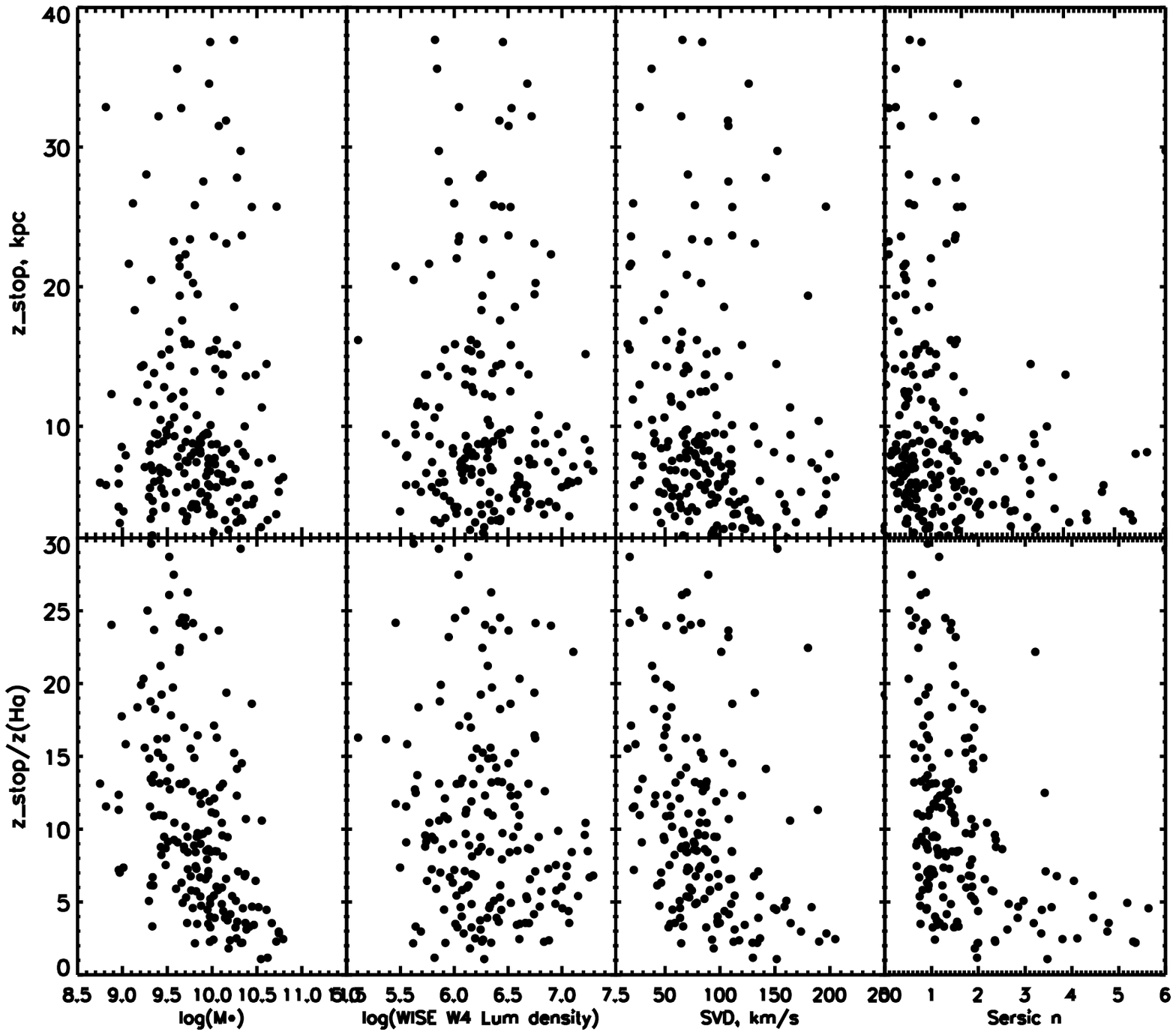}
\caption{The absolute (upper) and normalized (lower) stop altitude 
versus the galactic stellar mass, W4 luminosity density, central 
stellar velocity dispersion, and Sersic index. 
Note that the formally calculated correlation coefficients for all 
panels |CC| are less than 0.1, although some non-linear trends can be 
seen in the bottom panels. 
\label{fig6}}
\end{figure*}

While the lag indicates a gradient of rotation descending, we can 
formally calculate the distance in the vertical direction that
corresponds to a completely stopped rotation, $z_{stop}$, in the 
linear approximation. 
Introducing a spatial scale allows us to eliminate the size
effect in the lag trends. We define the stop-altitude
as $z_{stop} = V_{max} / lag$, where $V_{max}$ is the maximum 
rotation velocity. The latter is estimated in our modelling in \S.3.1.
In Figure~\ref{fig6} we consider trends of $z_{stop}$, and also
of the stop altitude normalized by the scale height of ionized gas
estimated as in B17. Similar to Figures~\ref{fig2}-\ref{fig4},
the absolute and normalized stop-altitude does not show any dependence
on the star formation rate or star formation density. At the same time
we notice that normalized stop-altitude is significantly shorter for massive
galaxies, galaxies with high velocity dispersion and with high
Sersic index. 

\subsection{Radial Gradients of Lags}

Following \citet{levy19}, we consider possible radial variations of lags
in the galaxies. We modified the pipeline described in \S.2.1 and 
added a possibility of a radial linear variation of the lags. 
\citet{levy19} used CALIFA data obtained for larger than MaNGA 
galaxies. To ensure that MaNGA data provide sufficient radial
resolution, we removed several galaxies in which the MaNGA IFU
covers the very central part only. Next, we compared the maximum
radial extent of our objects with the resolution element that 
corresponds to 2.5 arcsec resolution in the final MaNGA maps \citep{law16}. 
It turned out that only a few galaxies in the sample have less than 4 resolution
elements coverage from the centre to the edge, while most of 
selected galaxies have a better coverage. We considered the
number of resolution elements per MaNGA data radial extent 
as a parameter and explored whether the further trends of
radial variations of parameters discussed below depend on the
parameter. We do not find any dependence on the radial resolution
and conclude that MaNGA radial resolution is sufficient to explore 
the radial gradients of lags. 

\begin{figure}
\includegraphics[width=8.5cm]{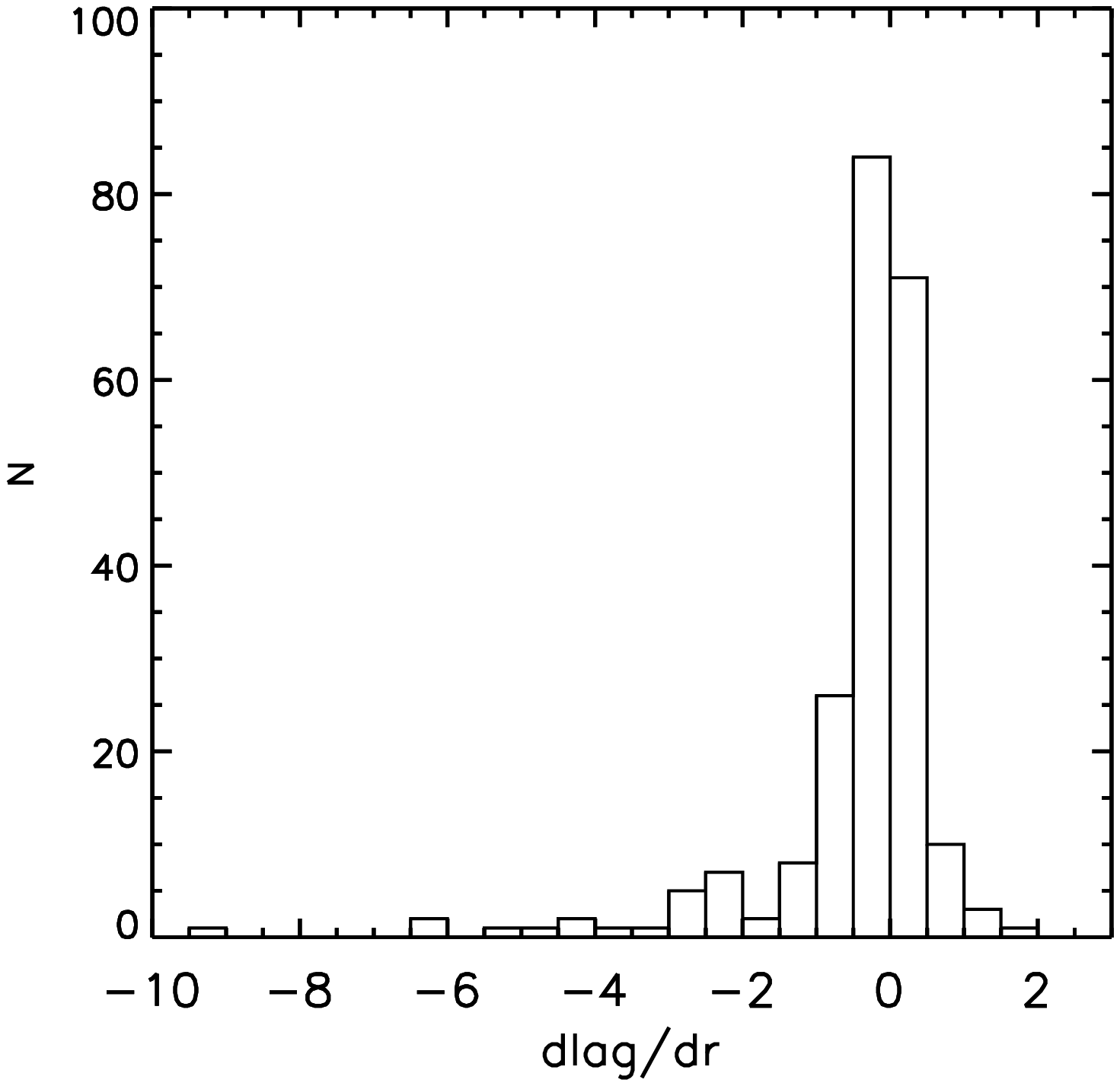}
\caption{The histogram distribution of the radial gradient of lags. 
While most of the galaxies show values close to zero, the distribution
has a "tail" towards the negative values. The tail (dlag/dr $<$ -2 \kms kpc$^{-2}$)
contains 8 \% of all galaxies. The 2 \kms kpc$^{-2}$ corresponds to the 
right limit of the quasi-gaussian centered at the zero value.
\label{fig6a}}
\end{figure}

\begin{figure}
\includegraphics[width=8.5cm]{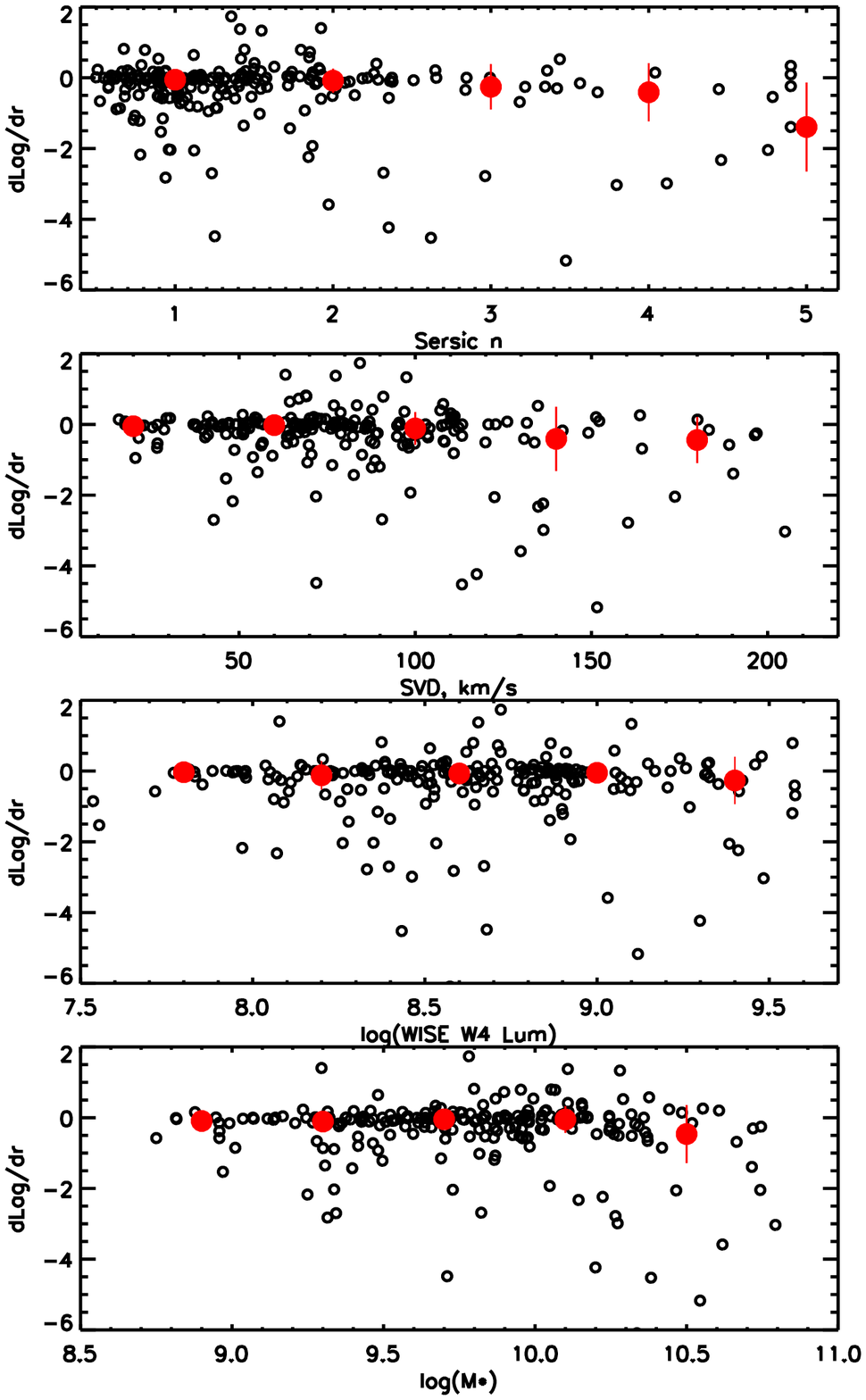}
\caption{The lag radial gradients in individual galaxies versus their
stellar mass, W4 luminosity, stellar velocity dispersion and Sersic index
(black circles). The red bullets designate the slipping median values
with 1.48 MAD for the error bars. 
\label{fig7}}
\end{figure}

The distribution of the radial variation of lags is shown in
Figure~\ref{fig6a}. Most of the galaxies indicate modest
radial lag gradients, although a few of them show significant
negative gradients. We explore the radial lag gradient trends
in Figure~\ref{fig7}. The black circles show the measurements in the 
galaxies, while the red bullets designate the slipping median and 
median absolute deviation (MAD) uncertainties in the sample. It is seen that the radial lag
gradient is essentially zero for all possible galaxy parameters. 
The gradient bias towards its negative values for massive galaxies
with large Sersic index is caused by small statistics in corresponding 
regions of parameter space. 

%According to \citet{levy19}, the radial lag gradient is an indicator
%of the lag origin.
%A negative gradient in the vertical lag with galactocentric radius might be expected in 
%galactic fountain models due to conservation of angular momentum
%\citep{shapiro76,bregman80, fraternali07,fraternali08,marinacci10,marinacci11}.
%A uniform infall of cicrumgalactic gas \citep{binney05} or a cylindrical accretion
%from the galactic corona \citep{kaufmann06} is not expected to cause a radial dependence of 
%the lag for the assumption of the accretion uniformity.
%In the picture of gas accretion with inflow considered by \citet{combes14}, 
%the angular momentum for the gas is redistributed in the disc such that 
%a positive radial gradient in the lag would be expected.

According to \citet{levy19}, the radial lag gradient is an indicator
of the lag origin.
The negative dlag/dr anticipates a larger lag at the central parts of  
galaxies than at its periphery due to the ejected gas momentum conservation
\citep{shapiro76,bregman80, fraternali07,fraternali08,marinacci10,marinacci11}
A uniform infall of cicrumgalactic gas \citep{binney05} or a cylindrical accretion
from the galactic corona \citep{kaufmann06} should not cause any lag gradients 
for the assumption of the accretion uniformity. The gas accretion from
the CGM at low inclination angles to the galactic midplane was studied 
by simulations \citep{stewart11} and was observed in some 
galaxies \citep{lehner13}. The gas accretion via the in-plane inflow 
redistributes the gas momentum in the disc, creates larger lags at the galactic
periphery, and thus causes positive lag radial 
gradients \citep[see also][]{putman12,combes14}.

\subsection{The Lag asymmetry between different sides of the galaxies}

\begin{figure}
\includegraphics[width=8.5cm]{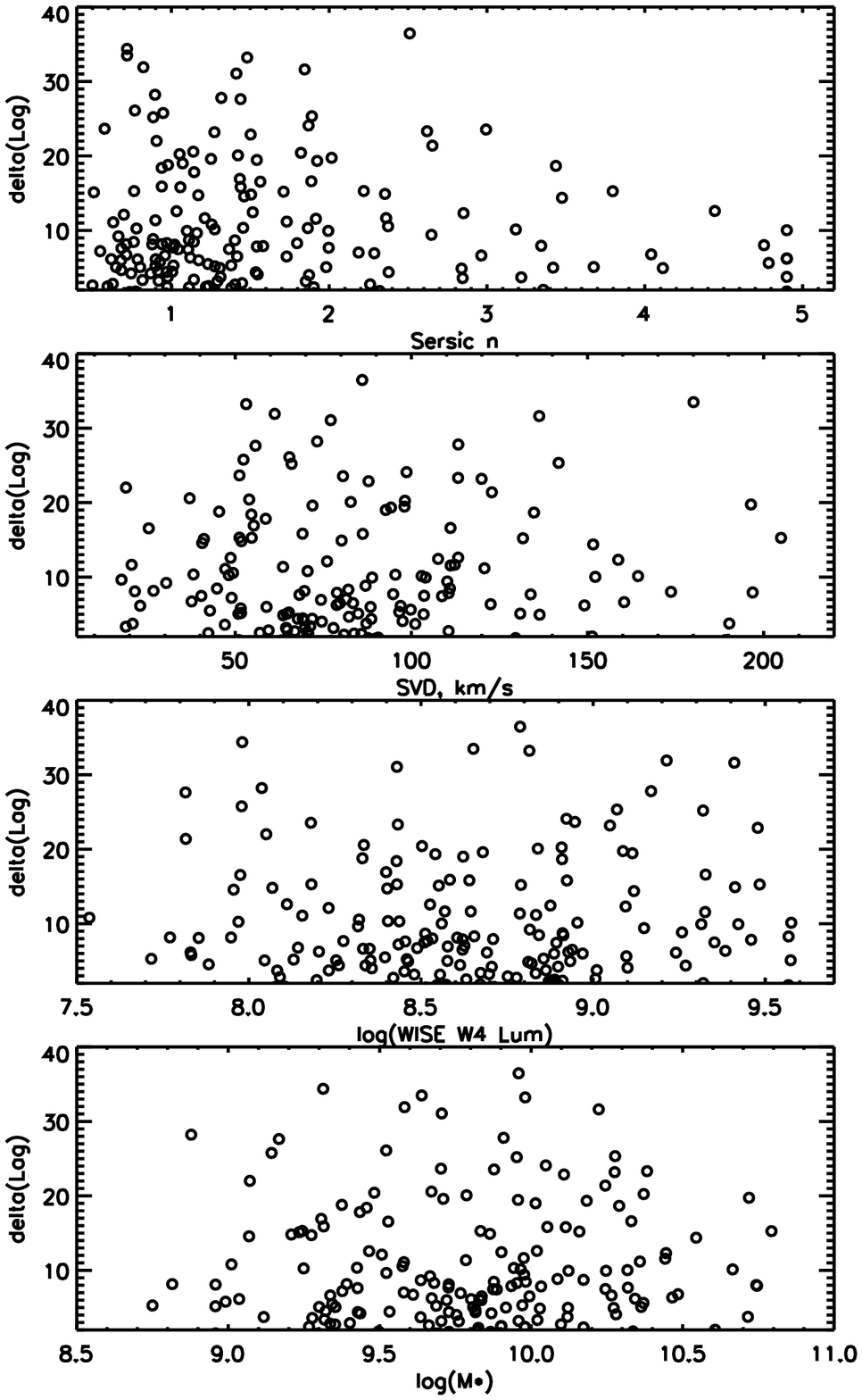}
\caption{The absolute difference $\Delta lag$ between the lag estimated for
the lower and upper parts of the galaxies versus the general galactic properties:
stellar mass, W4 luminosity, stellar velocity dispersion, and Sersic index. 
The formally calculated correlation coefficients |CC| for all panels above
are less than 0.1.
\label{fig8}}
\end{figure}

\begin{figure}
\includegraphics[width=8.5cm]{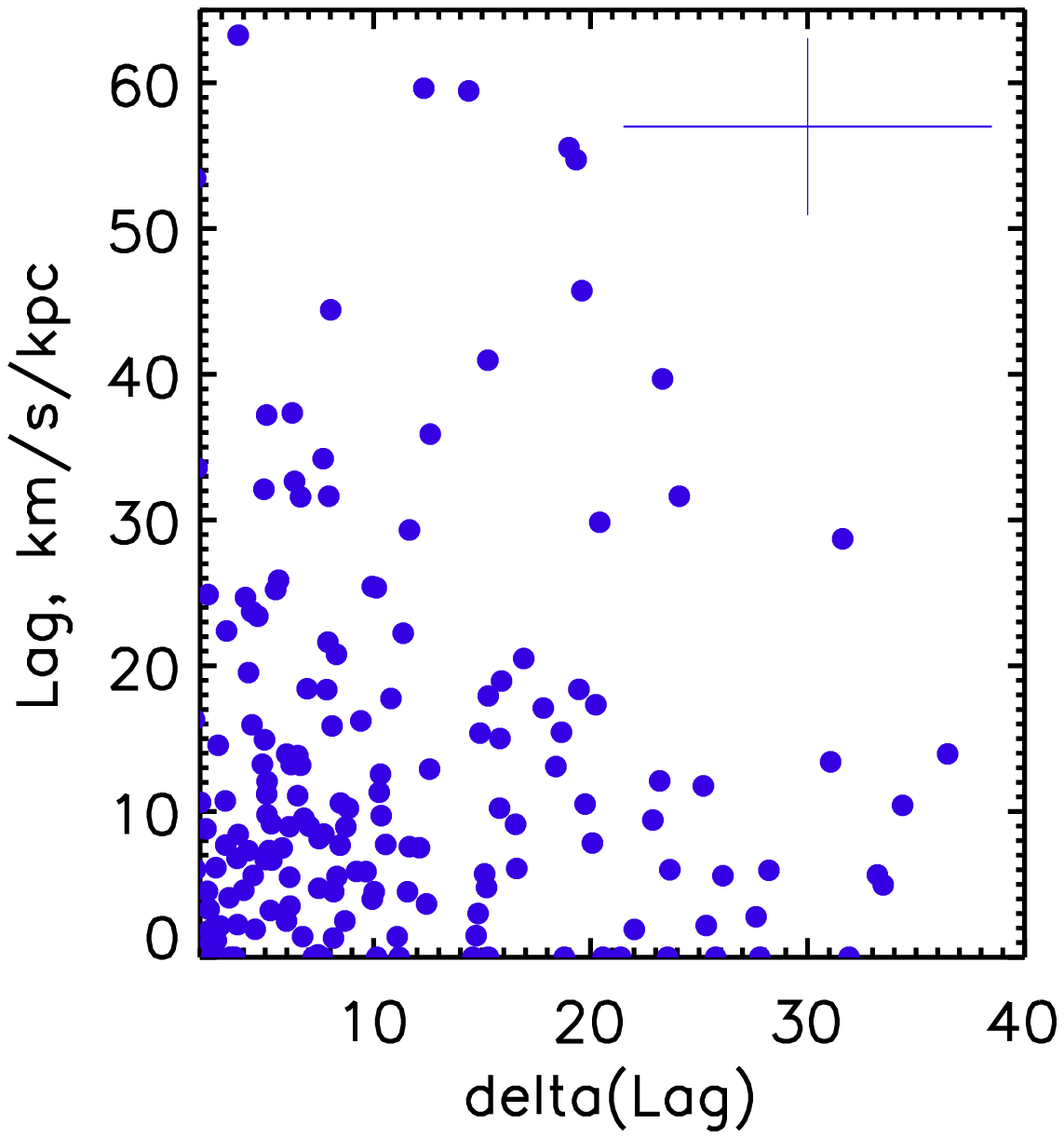}
\caption{Comparison of the lag and $\Delta lag$. The typical error bars 
are shown in the upper right corner of the plot.
\label{fig9}}
\end{figure}

Interaction of galactic gas with circumgalactic medium (CGM) may be affected
by non-uniformities of the CGM. Interactions with nearby satellites can 
also be a factor that affects galactic gas on the opposite sides from
the galactic midplane \citep{ghosh21} . Thus, asymmetric interactions may 
lead to different value of lag estimated above and below the galactic 
midplane. We use the same approach described in \S~3.1  and evaluated 
the same models for the "upper" and "lower" parts of the edge-on galaxies 
separately. Then we find the $\Delta$(lag) as the absolute difference 
between the lags on the two opposite halves of galaxies (with respect to the midplane). 
Figure~\ref{fig8} shows the $\Delta$(lag) as a function of stellar mass, 
stellar velocity dispersion, star formation rate and Sersic index. 
We cannot claim any clear linear trends.
We notice the lack of large $\Delta$(lag) values for galaxies with the Sersic n greater
than 3, although this area of the parameter space has a small number
of data points. 
Figure~\ref{fig9} shows that there is no correlation between the 
lags and $\Delta$(lag).

\subsection{Environment of Galaxies and the Lag Variation}

Since the lag origin is connected to the CGM and the environment of 
galaxies, we expect that the presence of some environmental structures
may correlate with strong rotation velocity lags. We were able to find images
of all our galaxies with regular lags in the LEGACY survey sky atlas
\citep{dey19}. We visually inspected the outskirts of each galaxy and searched for
such elements as low surface brightness shells, loops, bridges, streams,
extended spiral arms,
envelopes common with other galaxies, and also noticeable asymmetries of 
external isophotes - in a manner it was classified by \citep{atkinson13}.
In addition, we visually inspected extended areas around the galaxies and counted
possible satellites around them. We distinguished between small 
(non-stellar objects without noticeable structure), 
medium (with smaller size than the studied galaxies), 
large (comparable to the main galaxy in size), and great satellites
(larger than the main object). Since we have no information about redshifts of
the satellites in most cases, we expect that many small satellites 
are projected objects. At the same time, the medium, and especially
large and great satellites have a good chance to be at the same distance as 
our galaxies. 

We compared the number of the medium and larger sizes satellites (according 
to the classification above) for the galaxies in our sample
and did not find significant difference between the objects with large and 
small lags. 

\begin{figure}
\includegraphics[width=8.5cm]{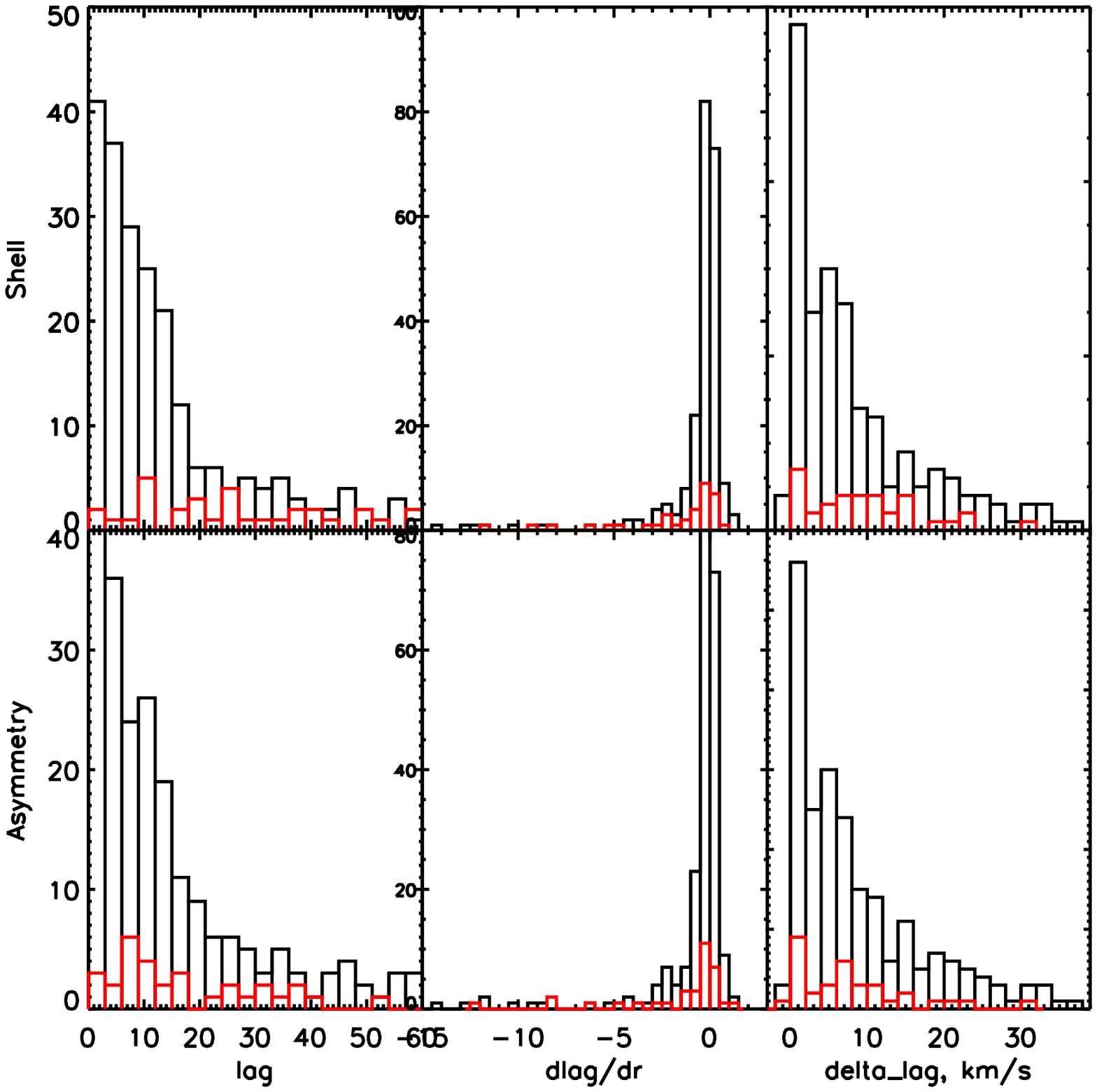}
\caption{The histogram distributions of the lag, lag radial gradient dlag/dr and 
$\Delta lag$ for the galaxies with (red) and without (black) shells 
(upper panels) and asymmetry of external isophotes (lower panels). 
The probability of similarities of the two samples is p=0.004, 0.018, and 0.73
for the upper panels, and p=0.130, 0.130, and 0.84 for the lower panels, see text.
\label{fig10}}
\end{figure}

We plot histograms of the lag, lag gradient and lag asymmetry ($\Delta lag$) distributions
for galaxies with and without two most commonly observed faint features
around them: low surface brightness shells and noticeable asymmetry of their 
external isophotes, see Figure~\ref{fig10}. 
The absence (black line) or presence (red line) of shells (upper panels) and 
outer asymmetries (lower panels) make some histograms look different.
We compare the distributions via the two-sample Kolmogorov-Smirnov (K-S)
test. The lags show large difference between the samples (probability of the similarity
is p=0.004 and 0.130 for the shells and asymmetry, respectively), 
as well as the lag gradient (p=0.018 and 0.130).
The lag asymmetry demonstrates a significant similarity for the low surface brightness 
shells and asymmetry with probabilities p=0.73 and 0.84, respectively.

We classified all identified environmental features around the galaxies by 
young, intermediate age and old, based on the stage of parental interaction.
The structures are called young if they are typical for the starting phase of
galactic interaction. We consider bridges, tight group membership, distortion due
to an evident interaction, streams, and tidal structures between 
interacting galaxies as young structures. On the opposite side there are old structures 
that are smooth and extended in space, e.g. shells, extended loops, inner and outer rings.
Formation of those structures requires more than one turn of a satellite around 
its host galaxy. 
We consider unusual dust pattern seen in the objects and arcs as structures of 
intermediate age. 

\begin{figure}
\includegraphics[width=8.5cm]{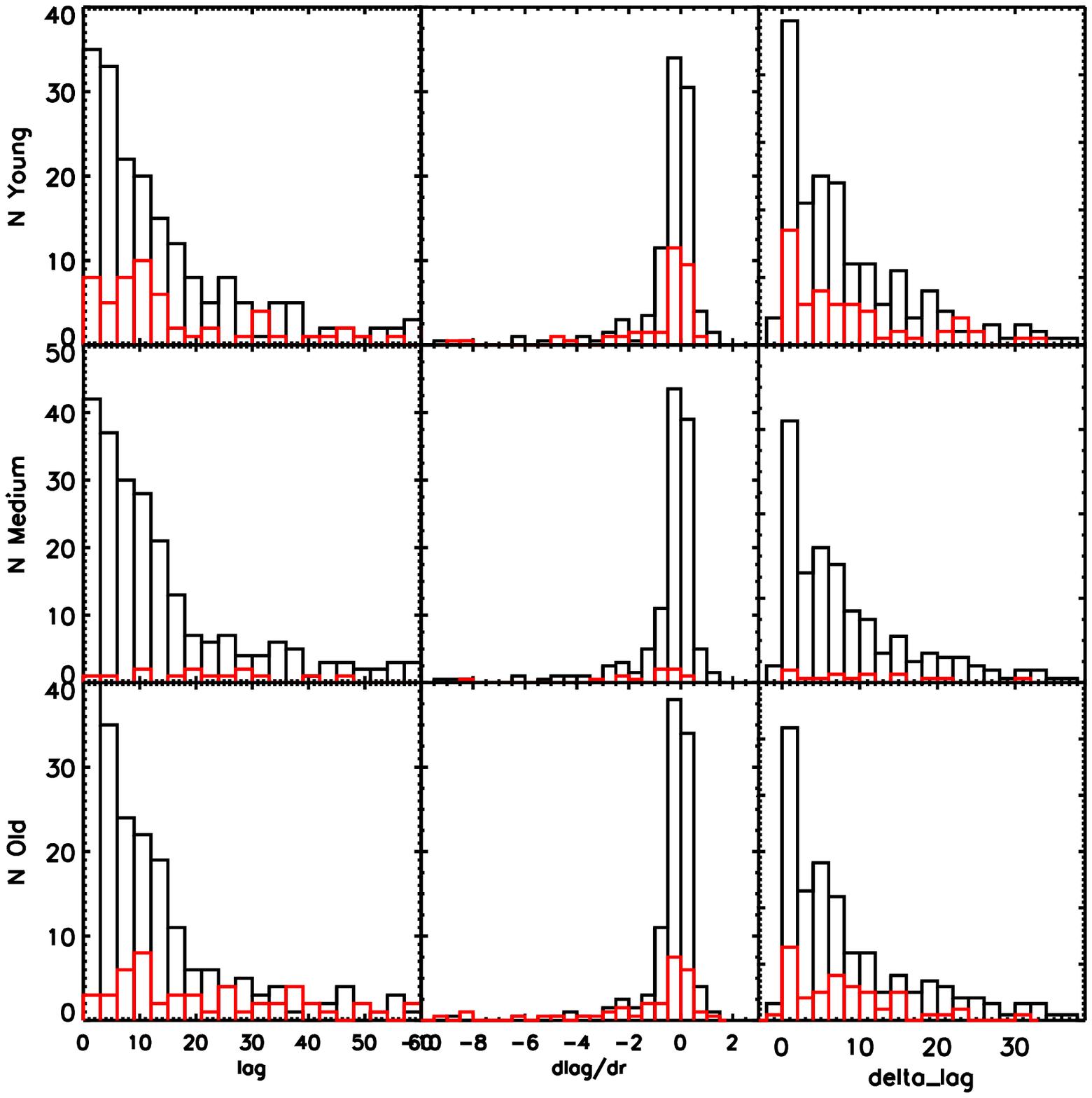}
\caption{The histogram distributions of the lag, lag radial gradient dlag/dr and 
$\Delta lag$ for the galaxies with (red) and without (black) 
Young (upper panels), Intermediate (middle panels) or Old (lower panels)
low surface brightness structures around them.
The probability of similarities of the two samples is p=0.72, 0.31, and 0.96
for the upper row, p=0.02, 0.09, and 0.33 for the middle row, 
and p=0.07, 0.01, and 0.93 for the lower panels, see text.
\label{fig11}}
\end{figure}

Figure~\ref{fig11} shows histogram distributions of lag, radial gradient of lag
and the lag asymmetry $\Delta lag$ for the young, intermediate
and old structures. The presence of the structures is designated with the 
red lines, while the absence of them with the black.
The lag distributions for the objects with young faint structures look
rather similar for the K-S test (p=0.72), while the intermediate
and old structures make difference (p=0.02 ans 0.07, respectively).
The lag gradients show some similarity for the young structures (p=0.31)
and difference for the intermediate and old (p=0.09 and 0.01, respectively).
The lag asymmetry looks very similar for the young and old features (p=0.96
and 0.93, respectively), while the histograms for intermediate age are similar 
with p=0.33.

%For ps_16, Fig 10
%More ENV Statistics KS
%lag, sh:    0.00392341
%lag, ao:      0.130149
%dlagdr, sh:     0.0181913
%dlagdr, ao:      0.130164
%dll, sh:      0.723661
%dll, ao:      0.837292
%For ps_17, Fig 11
%More ENV Statistics KS
%lag, YNG:      0.717634
%lag, MED:     0.0203821
%lag, OLD:     0.0725082
%dlagdr, YNG:      0.314597
%dlagdr, MED:     0.0906499
%dlagdr, OLD:     0.0111767
%dll, YNG:      0.956992
%dll, MED:      0.330381
%dll, OLD:      0.926072

\section{Results and Discussion}

We confirm general trends of the lag and lag gradient found by \citet{bizyaev17} and
\citet{levy19} with our significantly larger sample. We don't find 
any correlation between the star formation activity (defined
by the integrated rate or per unit surface area) in galaxies 
and their gas rotation lag. The lag value correlates with galactic
stellar mass, rotation curve amplitude and central stellar velocity dispersion. 
The stop altitude introduced in this paper is seen as a proxy for the
demarcation between the internal and external gas from the
standpoint of its kinematics.
When normalized by the scaleheight of ionized gas (Figure~\ref{fig6}), the stop altitude
suggests that low mass galaxies (stellar mass less than 10$^{10} \, M_{\odot}$) 
with low Sersic index and stellar velocity
dispersion posses a wider "zone of influence" with respect to massive
galaxies with a large contribution of spheroidal component. From scaling
up size with mass in galaxies and by assuming the same density of
surrounding circumgalactic gas, we could expect to see more extended "zone
of influence" of internal gas in massive galaxies with respect to 
low mass objects. Our Figure~\ref{fig6} shows the opposite. 
This, in turn, points at a difference in the properties of gas surrounding
low- and high-mass galaxies and
suggests that interaction between the internal and external
gas plays a critical role in the formation of lags - at least in a 
majority of the galaxies in our MaNGA edge-on sample. 

At the same time, the lag radial gradient is found negative in a small fraction 
of all galaxies, as shown in Figure~\ref{fig6a}. According to \citet{levy19}
and references therein, the negative radial gradients point at the internal
origin of lags and their connection to galactic fountains. We see noticeable
negative gradients in only 8 \% of our galaxies, see Figure~\ref{fig6a}. 
There is no particular mass, Sersic index, or star formation
rate range correlated with the radial lags.
Our results do not indicate that the lags caused by galactic fountains 
are ubiquitous or can be predicted from some general galactic properties. 
Instead, most of the galaxies demonstrate negligible radial gradients of
lags, consistent with a gas accretion scenario \citep{levy19}.
Note that the vertical lags generally have been found too large 
to be explained purely with galactic fountain models and that 
accretion has been suggested \citep{fraternali07,fraternali08,marinacci10}.

The lag asymmetry, defined as a difference between the lag determined 
on both sides from the galactic midplane independently, does not show 
a significant correlation with general galactic properties (Figure~\ref{fig8}).

Our study of galactic environment shows that some stellar low surface brightness
structures can be seen more often in galaxies with larger lags. According
to Figure~\ref{fig9}, low surface brightness shells and outer isophote
asymmetry are more frequently observed in galaxies with large lags and 
with significant negative radial lag gradients. At the same time, these 
faint structures do not depend on the lag asymmetry. 

When interaction-borne environmental structures are subdivided by young, 
intermediate, and old based on their typical age of interaction coarsely
determined from the structure extension and geometry, the lag 
and radial lag gradient histogram distributions are significantly different
between the samples with and without the old environmental structures
around. The sample with the old structures around is biased towards 
the objects with large lags and negative lag gradients, see Figure~\ref{fig11}. 
The young structures around the galaxies do not affect their
lag and lag radial gradient distributions. We cautiously avoid from making 
conclusions for the intermediate age structures because of poor statistics
in this case. Our results suggest that we observe significant coincidence
of old environmental structures and the high frequency of large lags and negative 
radial lag gradients in galaxies. Similar to the case of galactic fountains scenario
when the expelled gas returns back to its host galaxy, interaction with 
galactic environment, which is often accompanied by the gas infall, causes 
large radial lag gradients. As it can be seen in Figure~\ref{fig11}, the gas infall
that took place in the past increases the amplitude of lags and causes negative
lag radial gradients. At the same time, for most of our galaxies we do not
observe significant radial gradients of lags, and also we don't find any connection between
the lag asymmetry and past interaction events. This allows us to reject the gas accretion
caused by interaction with satellites scenario for most of the galaxies. In turn, it favours 
the gas accretion from the CGM as the main source of rotation velocity lags -
same conclusion as reported in a recent, similar study by \citet{levy19}, 
but with a much larger sample.

\section{Conclusions}

We considered the largest sample of \Neon edge-on galaxies observed up to date with 
integral field units by the MaNGA survey. Among them, \Ng galaxies
show regular decreasing of their rotation velocity on their ionized
gas velocity fields, or simply lag. 

We consider how the lags and the stop altitude (defined as the distance at which 
the gas rotation should stop, in the linear approximation), depend on general properties 
of galaxies. We do not find any correlation of the lags or stop altitude
with the star formation activity in the galaxies. Instead, mild correlations
of the stop altitude found in the paper suggest that low mass galaxies with 
low Sersic index and with low stellar velocity dispersion - the objects that should have a 
high specific fraction of gas with respect to stars, -  posses wide
"zone of influence" in the extragalactic gas surrounding them. In contrast, 
massive galaxies with large contribution of spherical component experience more
significant effects from the extragalactic gas which imprints in kinematics
of their extraplanar gas.

We estimated the radial gradients of lags and find them flat 
for most of the galaxies in our sample, which suggests connection
between the lags and accretion from the intergalactic media. 
A small subsample of galaxies with negative radial gradients of lag
have enhanced fraction of objects with aged low surface brightness 
structures around them (e.g. faint shells), which indicates that 
noticeable satellite accretion events in the past affected the extraplanar 
gas kinematics and might have caused negative lag gradients. 

We considered the lag asymmetry between the "upper" and "lower"
parts of the galaxies with respect to galactic midplanes, and found
neither correlation of it with galactic parameters nor its connection
to the galactic environment. 

Based on the high frequency of zero radial gradient of lag
in a combination with low lag asymmetry in most of our galaxies, 
we conclude that isotropic accretion of gas from the CGM plays 
a significant role in the formation of gas rotation velocity lags.

\section*{Acknowledgements}
We thank the anonymous referee for constructive feedback that improved the paper.
This work is partially supported by the National Science Foundation under 
Grant No. AST-1615594 to RAMW.
Y. C acknowledges support from the National Natural Science Foundation 
of China (NSFC grants 11573013, 11733002, 11922302).
RAR acknowledges  partial financial support from CNPq and FAPERGS.
The project is partly supported by RSCF grant 19-12-00145.
SDSS-IV acknowledges support and resources from the Center for
High-Performance Computing at the University of Utah.  The SDSS web site
is www.sdss.org.

SDSS-IV is managed by the Astrophysical Research Consortium for the
Participating Institutions of the SDSS Collaboration including the
Brazilian Participation Group, the Carnegie Institution for Science,
Carnegie Mellon University, the Chilean Participation Group, the French
Participation Group, Harvard-Smithsonian Center for Astrophysics,
Instituto de Astrof\'isica de Canarias, The Johns Hopkins University,
Kavli Institute for the Physics and Mathematics of the Universe (IPMU) /
University of Tokyo, Lawrence Berkeley National Laboratory, Leibniz
Institut f\"ur Astrophysik Potsdam (AIP), Max-Planck-Institut f\"ur
Astronomie (MPIA Heidelberg), Max-Planck-Institut f\"ur Astrophysik (MPA
Garching), Max-Planck-Institut f\"ur Extraterrestrische Physik (MPE),
National Astronomical Observatory of China, New Mexico State University,
New York University, University of Notre Dame, Observatrio Nacional /
MCTI, The Ohio State University, Pennsylvania State University, Shanghai
Astronomical Observatory, United Kingdom Participation Group, Universidad
Nacional Aut\'onoma de M\'exico, University of Arizona, University of
Colorado Boulder, University of Oxford, University of Portsmouth,
University of Utah, University of Virginia, University of Washington,
University of Wisconsin, Vanderbilt University, and Yale University.

\section*{DATA AVAILABILITY}
This work makes use of SDSS/MaNGA project data publicly available at
https://www.sdss.org/dr17/data\_access/.

{}

\bsp    % typesetting comment
\label{lastpage}

\end{document}